\documentclass[runningheads]{llncs}
\usepackage[T1]{fontenc}
\usepackage{graphicx}
\usepackage{hyperref}
\usepackage{color}

\usepackage{listings}

\lstdefinestyle{SingularityStyle}{
  upquote=true,
  breaklines=true,
  breakatwhitespace=true,
  basicstyle=\fontsize{6}{8}\selectfont\ttfamily,
  frame=tblr,
  aboveskip=10pt,
  belowskip=10pt,
  showstringspaces=false,
  linewidth=1.0\linewidth,
  xleftmargin=0.0\linewidth,
  keywordstyle=\bfseries,
  morekeywords={Bootstrap,From,post,files,environment,runscript},
}

\lstdefinestyle{TerminalStyle}{
  upquote=true,
  breaklines=true,
  breakatwhitespace=true,
  basicstyle=\fontsize{6}{8}\selectfont\ttfamily,
  frame=tblr,
  aboveskip=10pt,
  belowskip=10pt,
  showstringspaces=false,
  linewidth=1.0\linewidth,
  xleftmargin=0.0\linewidth,
  keywordstyle=\bfseries,
  morekeywords={\$},
  escapechar=@
}

\definecolor{mygray}{rgb}{0.4,0.4,0.4}
\definecolor{mygreen}{rgb}{0,0.8,0.6}
\definecolor{myorange}{rgb}{1.0,0.4,0}

\lstdefinestyle{C_Style}{
  language=C,
  upquote=true,
  breaklines=true,
  breakatwhitespace=true,
  basicstyle=\fontsize{6}{8}\selectfont\ttfamily,
  frame=tblr,
  aboveskip=0pt,
  belowskip=0pt,
  showstringspaces=false,
  linewidth=1.0\linewidth,
  xleftmargin=0.0\linewidth,
  emph={MPI_CALL,printf,exit,fprintf},
  emphstyle={\color{blue}},
  commentstyle=\color{mygray},
  stringstyle=\color{myorange},
  keywordstyle=\color{mygreen},
  tabsize=2
}
\begin{document}
\title{Enabling Message Passing Interface Containers on the LUMI Supercomputer}
%
%
\author{Alfio Lazzaro\orcidID{0000-0003-4256-8270}
}
\authorrunning{A. Lazzaro}
%
\institute{HPE HPC/AI EMEA Research Lab, \\
Innere Margarethenstrasse 5, 4051 Basel, Switzerland \\
\email{alfio.lazzaro@hpe.com}\\
\url{https://www.hpe.com/emea_europe/en/compute/hpc/emea-research-lab.html}
}
\maketitle              
\begin{abstract}
Containers represent a convenient way of packing applications with dependencies for easy user-level installation and productivity. When running on supercomputers, it becomes crucial to optimize the containers to exploit the performance optimizations provided by the system vendors. In this paper, we discuss an approach we have developed for deploying containerized applications on the LUMI supercomputer, specifically for running applications based on Message Passing Interface (MPI) parallelization. We show how users can build and run containers and get the expected performance. The proposed MPI containers can be provided on LUMI so that users can use them as base images. Although we only refer to the LUMI supercomputer, similar concepts can be applied to the case of other supercomputers.
\end{abstract}
\section{Introduction}

Containers represent a convenient way of packaging applications with dependencies for easy user-level installation and productivity. They are used to package entire scientific workflows, software, libraries, and even data, solving the problem of making the software run reliably when moved from one computing environment to another.
They provide a simple way of sharing scientific applications and reproducing research on either cloud or High Performance Computing (HPC) systems. The \emph{de-facto} standard technology is Docker, which is widely used in cloud environments~\cite{merkel2014docker}. However, mainly due to security concerns, Docker containers are not directly suitable for running on HPC systems~\cite{securityhpc}. For this reason, optimized technologies have been developed for running on HPC systems such as Shifter~\cite{gerhardt2017shifter}, Charliecloud~\cite{priedhorsky2017charliecloud}, Singularity~\cite{kurtzer2017singularity}, Sarus~\cite{sarus}, Apptainer~\cite{osti_1886029}, or adapted for use on HPC, like Podman~\cite{heon_2021_4735634}.

In this paper, we will discuss the suggested procedures for deploying HPC application containers on the LUMI supercomputer~\footnote{\url{https://www.lumi-supercomputer.eu/}.}, specifically for running applications based on Message Passing Interface (MPI) parallelization. 
LUMI provides the Singularity runtime for running containers. The Singularity privilege model relies on SUID and non-privileged user namespaces to launch containers as child processes, thus allowing non-root users to create and launch containers safely. However, on LUMI user namespaces are disabled because of possible security risks concerns~\cite{cve_userns}. As a result of that, the Singularity fakeroot feature is not available and the users cannot build their containers on LUMI~\cite{fakeroot}, which is a desirable possibility for fast optimizing and testing of containers directly on the system where they will be eventually run in production. To address these cases, Singularity (since SingularityCE 3.11) has introduced the possibility to build containers via \texttt{proot}, with some limitations~\footnote{\url{https://docs.sylabs.io/guides/3.11/user-guide/build_a_container.html\#unprivilged-proot-builds}.}. We will describe how this technique has been deployed on LUMI in~Section~\ref{sec:building}.

MPI is extensively used by HPC applications to implement communications across parallel processes that are running on computed nodes connected by a high-speed interconnect. LUMI is an HPE Cray EX system that features the HPE Slingshot interconnect with 200 Gbps HPE Slingshot Cassini (CXI) network adapters~\cite{slingshot}. The HPE Cray MPI 
is available on the system~\cite{craympich}. This is a proprietary and optimized implementation of the MPI specification, based on the open-source MPICH CH4 implementation from the Argonne National Laboratory~\cite{mpich}. HPE Cray MPI relies on the OFI Libfabric interface~\cite{libfabric} with the CXI network provider to leverage the broad set of hardware capabilities offered by the HPE Slingshot network. Therefore, it is crucial to use the HPE Cray MPI to get high performance when running MPI applications in containers. This part will be described in Section~\ref{sec:running}, where we include some testing procedures and performance results.

\section{Building Containers on LUMI}
\label{sec:building}

The Singularity version installed on LUMI is \texttt{singularity-ce}\linebreak
\texttt{version 3.11.4-1}. This version can build \emph{rootless} (i.e. as a non-root user) containers when \texttt{proot} is available on the system \texttt{PATH}. On LUMI, this can be achieved via the module \texttt{systools}~\footnote{\url{https://lumi-supercomputer.github.io/LUMI-EasyBuild-docs/s/systools/}.}, provided by the LUMI User Support Team (LUST). An example of how to access the command is shown in~Fig.~\ref{fig:proot}. Then, Singularity containers can be built via the usual \texttt{singularity build} command.

\begin{figure}[t]
\begin{lstlisting}[style=TerminalStyle]
$ module load LUMI/23.09 partition/C systools
$ proot --version
 _____ _____              ___
|  __ \  __ \_____  _____|   |_
|   __/     /  _  \/  _  \    _|
|__|  |__|__\_____/\_____/\____| 5.4.0

built-in accelerators: process_vm = yes, seccomp_filter = yes

Visit https://proot-me.github.io for help, bug reports, suggestions, patches, ...
Copyright (C) 2023 PRoot Developers, licensed under GPL v2 or later.
\end{lstlisting}
\caption{An example of how to access the \texttt{proot} command (version 5.4.0) on LUMI via Lmod modules, provided by LUST. The command itself is provided by the \texttt{systools} module, which is in this example part of the software hierarchy provided by the modules \texttt{LUMI/23.09} and \texttt{partition/C}. }
\label{fig:proot}
\end{figure}

For what concerns the description in this paper, we are interested in building containers with MPI. Nevertheless, similar concepts can be considered for the building of any container. For running a Singularity container with MPI on LUMI we use the so-called \emph{Hybrid} model~\footnote{\url{https://docs.sylabs.io/guides/3.11/user-guide/mpi.html\#hybrid-model}.}~\cite{hybrid_mode}. In this approach, we install a version of MPI inside the container to build it, but then we rely on the HPE Cray MPI available on LUMI during its execution. The prerequisite is that the MPI in the container must be compatible with the HPE Cray MPI. The HPE Cray MPI source code is not publicly available, however, HPE maintains the compatibility with the open-source  MPICH implementation, namely MPICH version 3.4a2~\cite{craympich}. A minimal example of Singularity definition file for a container with MPICH 3.4a2 is shown in Fig.~\ref{fig:defmpich}.

\begin{figure}[t]
\begin{lstlisting}[style=SingularityStyle]
Bootstrap: docker
From: ubuntu:24.04

%post -c /bin/bash
  apt-get update && apt-get -y upgrade --no-install-recommends
  apt-get -y install --no-install-recommends \
          build-essential wget file ca-certificates \
          gfortran

  # Cleanup
  apt-get autoremove -y
  apt-get clean && rm -rf /var/lib/apt/lists/*

  # Installation dir
  export INSTALL_DIR=/container
  mkdir -p ${INSTALL_DIR}

  VER=3.4a2
  wget -q http://www.mpich.org/static/downloads/${VER}/mpich-${VER}.tar.gz
  tar xf mpich-${VER}.tar.gz && rm mpich-${VER}.tar.gz
  cd mpich-${VER}
  sed -i 's/libmpi_so_version="0:0:0"/libmpi_so_version="12:0:0"/g' configure
  FFLAGS='-fallow-argument-mismatch' \
    ./configure --prefix=${INSTALL_DIR}/mpi --disable-static \
                --disable-rpath --disable-wrapper-rpath \
                --enable-fast=all,O3 --with-device=ch3 \
                --mandir=/usr/share/man > /dev/null
  make -j$(getconf _NPROCESSORS_ONLN) install > /dev/null
  cd .. && rm -rf mpich-${VER}
  echo "export PATH=${INSTALL_DIR}/mpi/bin:\$PATH" >> ${SINGULARITY_ENVIRONMENT}
  echo "export LD_LIBRARY_PATH=\${LD_LIBRARY_PATH}:${INSTALL_DIR}/mpi/lib" >> \
    ${SINGULARITY_ENVIRONMENT}
\end{lstlisting}
\caption{Minimal example of Singularity definition file to build MPICH compatible with the HPE Cray MPI available on LUMI. We use the Ubuntu 24.04 base image from Docker Hub. Note that we are specifically disabling  the static libraries creation and the \texttt{rpath} of the shared libraries. 
Furthermore, we use the CH3 implementation, which doesn't require to install any network library interface, e.g. OFI Libfabric library. This allows binding the HPE Cray MPI when running the container.
}
\label{fig:defmpich}
\end{figure}

Although OpenMPI implementation~\cite{openmpi} is not compatible with HPE Cray MPI, there is still a way to run a container based on it on LUMI. This is particularly interesting since the OpenMPI implementation is quite popular as the default MPI implementation in most common Linux distributions. Our approach makes use of the HPE \texttt{MPIxlate} tool~\footnote{\url{https://cpe.ext.hpe.com/docs/mpt/mpixlate/index.html}}. \texttt{MPIxlate} enables applications compiled using an MPI library that is not binary compatible with HPE Cray MPI, to be run without recompilation on supported HPE Cray systems. It does transparent runtime translation of the Application Binary Interface (ABI) between supported combinations of source and target MPI shared library implementations, in our case between OpenMPI version 4.x and HPE Cray MPI. Note that the ABI translation for MPI applications and libraries written in Fortran is not supported. A minimal example of Singularity definition file for a container with OpenMPI 4.1.6 (from the Ubuntu 24.04 release) is shown in Fig.~\ref{fig:defopenmpi}.

\begin{figure}[t]
\begin{lstlisting}[style=SingularityStyle]
Bootstrap: docker
From: ubuntu:24.04

%post -c /bin/bash
  # fake some of the commands not available with proot
  for f in /usr/sbin/groupadd /usr/sbin/addgroup /bin/chgrp; do
      rm -rf $f
      ln -s /bin/true $f
  done

  apt-get update && apt-get -y upgrade --no-install-recommends
  apt-get -y install --no-install-recommends \
          build-essential wget libopenmpi-dev

  # Cleanup
  apt-get autoremove -y
  apt-get clean && rm -rf /var/lib/apt/lists/*
\end{lstlisting}
\caption{Minimal example of Singularity definition file with OpenMPI (version 4.1.6), based on the Ubuntu 24.04 base image from Docker Hub. The \texttt{openssl} package, a dependency of the \texttt{libopenmpi-dev}, requires to run some privileged commands (\texttt{groupadd, addgroup, chgrp}), for which the  \texttt{proot}'s emulation of the root user doesn't work. As a workaround, we \emph{fake} them with symbolic links to the \texttt{true} command, so that they will always return without errors, allowing the successful build of the container. A more systematic approach to fake privileged commands has been proposed in Charliecloud via the kernel's \texttt{seccomp(2)} system call filtering~\cite{charliecloud_seccomp}. 
}
\label{fig:defopenmpi}
\end{figure}

A different approach for running OpenMPI containers on HPE Cray supercomputers has been proposed in Ref.~\cite{cscs_libfabric}. The idea is to build the container with OpenMPI and OFI Libfabric support and then 
bind the LUMI OFI Libfabric library when executing the container, i.e. applying the hybrid mode at the level of the OFI Libfabric library. The drawbacks are that we don't make use of the optimized HPE Cray MPI and that we have to carefully install an OFI Libfabric library in the container compatible with the LUMI OFI Libfabric library, making the Singularity definition file more complex. For these reasons, we decided not to test this approach on LUMI, leaving its testing as future work.

So far, we have only discussed MPI implementations where the CPU buffers are used for communications. A separate discussion concerns the possibility of deploying containers with GPU-aware MPI functionalities, i.e. when the MPI implementation can send and receive GPU buffers directly, without having to first stage them in host memory. This is relevant for the LUMI GPU partition with AMD Mi250X GPUs. HPE Cray MPI supports GPU-aware MPI communications via the HPE proprietary GPU Transport Layer (GTL) library~\cite{craympich}. In particular, the GTL library has to be included during the linking phase of the applications within the containers, which will break the container portability.
Furthermore, it is important to note that MPICH provides native GPU support only since version 4.x, which is not ABI compatible with HPE Cray MPI, and in general AMD GPU support is only available through the UCX communication framework, which is not compatible with the OFI Libfabric framework. 
Enabling OpenMPI with native GPU-aware on HPE Cray EX systems has been presented in Ref.~\cite{openmpi_frontier}. Eventually, we will investigate how to extend our solutions to include GPU-aware MPI implementations as future work.

\section{Running Containers with MPI on LUMI}
\label{sec:running}

We use the MPICH container presented in Section~\ref{sec:building} as a base image for building an application container. For our description, we consider a small MPI example (see Fig.~\ref{fig:mpitest}) and the OSU Micro Benchmarks~\footnote{\url{https://mvapich.cse.ohio-state.edu/benchmarks/}.} as applications. In this way, we can test if MPI is properly running and its performance, respectively. An MPI application will crash with an error during the \texttt{MPI\_Init} execution call if MPI is not properly initialized during the batch execution. On LUMI the workload manager is Slurm. However, it can also happen that the application doesn't report any error at the init phase, pretending to run multiple identical sequential instances with an eventual crash at a later stage of the execution. To avoid such a case, we adopt the good practice of including a test during the \texttt{MPI\_Init} via a \texttt{LD\_PRELOAD} intercept of the call to make sure that the requested number of MPI ranks in Slurm is effectively the same of the running MPI ranks (following the principle of \emph{failing fast}). The source code of this test is shown in Fig.~\ref{fig:init_test}. 
We put everything in a Singularity definition file, presented in Fig.~\ref{fig:mpich_app}, which we build on LUMI.

\begin{figure}
\begin{lstlisting}[style=C_Style]
#include "mpi.h"
#include <stdio.h>
#include <stdlib.h>

#define MPI_CALL(func, args)                                                  \
  {                                                                           \
    int ierror = func args;                                                   \
    if (ierror != MPI_SUCCESS) {                                              \
      printf("\nMPI error: %s failed (%s::%d)\n", #func, __FILE__, __LINE__); \
      exit(EXIT_FAILURE);                                                     \
    }                                                                         \
  }

int main(int argc, char *argv[])
{
  int myrank = -1, nranks = -1, len = 0;
  char version[MPI_MAX_LIBRARY_VERSION_STRING];

  MPI_CALL(MPI_Init, (&argc,&argv));
  MPI_CALL(MPI_Comm_rank, (MPI_COMM_WORLD, &myrank));
  MPI_CALL(MPI_Comm_size, (MPI_COMM_WORLD, &nranks));
  if (myrank == 0) {
    printf("# ranks = %d (rank id = %d)\n\n", nranks, myrank);
    MPI_CALL(MPI_Get_library_version, (version, &len));
    printf("%s\n", version);
  }
  MPI_CALL(MPI_Finalize, ());

  return EXIT_SUCCESS;
}
\end{lstlisting}
\caption{MPI test example (\texttt{mpitest.c}) used to check that MPI is properly running. A correct output reports the corrected number of MPI ranks and the version of the MPI implementation library.
}
\label{fig:mpitest}
\end{figure}

\begin{figure}
\begin{lstlisting}[style=C_Style]
#define _GNU_SOURCE

#include <dlfcn.h>
#include <stdio.h>
#include "mpi.h"
#include <stdlib.h>
#include <string.h>
#include <stdbool.h>

#define MPI_CALL(func, args)                                            \
  *ierror = func args;                                                  \
  if (*ierror != MPI_SUCCESS)                                           \
    return;

void mpi_init_check(int* ierror) {
  if (*ierror==MPI_SUCCESS) {
    const char* senv = getenv("CHECK_MPI");
    if (senv!=NULL && (senv[0]!='0' || strlen(senv)>1)) {
      bool senv_verbose = (senv[0]=='2' && strlen(senv)==1);
      int myrank, nranks;
      MPI_CALL(MPI_Comm_rank, (MPI_COMM_WORLD, &myrank));
      MPI_CALL(MPI_Comm_size, (MPI_COMM_WORLD, &nranks));
      const char* slurm_ntasks_env = getenv("SLURM_NTASKS");
      if (slurm_ntasks_env!=NULL) {
        const int slurm_ntasks = atoi(slurm_ntasks_env);
        if (slurm_ntasks!=nranks) {
          fprintf(stderr, "ERROR: # MPI ranks %d != # Slurm tasks %d (rank id = %d)\n",
                  nranks, slurm_ntasks, myrank);
          fflush(0);
          MPI_Abort(MPI_COMM_WORLD, EXIT_FAILURE);
        }
        else if (senv_verbose) {
          fprintf(stderr, "INFO: # MPI ranks %d == # Slurm tasks %d (rank id = %d)\n",
                  nranks, slurm_ntasks, myrank);
          fflush(0);
          MPI_CALL(MPI_Barrier, (MPI_COMM_WORLD));
        }
      }
      else {
        fprintf(stderr, "WARNING: SLURM_NTASKS variable not declared (rank id = %d / # ranks = %d)!\n", myrank, nranks);
        fflush(0);
        MPI_CALL(MPI_Barrier, (MPI_COMM_WORLD));
      }
    }
  }
}

// C intercept call
int MPI_Init(int *argc, char ***argv) {
  int (*original_MPI_Init)(int *argc, char ***argv);
  original_MPI_Init = dlsym(RTLD_NEXT, "MPI_Init");
  int return_value = (*original_MPI_Init)(argc, argv);
  mpi_init_check(&return_value);
  return return_value;
}

// Fortran intercept call
void mpi_init_(int *ierror) {
  void (*original_mpi_init_)(int *ierror);
  original_mpi_init_ = dlsym(RTLD_NEXT, "mpi_init_");
  (*original_mpi_init_)(ierror);
  mpi_init_check(ierror);
}
\end{lstlisting}
\caption{\texttt{MPI\_Init} intercept call (C and Fortran APIs) to be used via \texttt{LD\_PRELOAD}. By default, the test is disabled. It can be enabled by setting the environment variable \texttt{CHECK\_MPI=1} (or \texttt{2} for a verbose mode). Then, the Slurm environment variable \texttt{SLURM\_NTASKS} is compared to the number of running MPI ranks and the execution aborts if the two values differ. The test can be part of the base MPI container.
}
\label{fig:init_test}
\end{figure}

The final step before running the container is to properly bind the host HPE Cray MPI into the Singularity container. This is achieved by setting the paths and libraries via the \texttt{SINGULARITY\_BIND} and \linebreak \texttt{SINGULARITYENV\_LD\_LIBRARY\_PATH} host environment variables, receptively. We extract the values by compiling and running the~\texttt{mpitest.c} (Fig.~\ref{fig:mpitest}) directly on LUMI, i.e. without containers, under the command \texttt{strace -f -e trace=\%file}~\footnote{\url{https://man7.org/linux/man-pages/man1/strace.1.html}.}. Then, a script parses the output to extract the values of  \texttt{SINGULARITY\_BIND} and \texttt{SINGULARITYENV\_LD\_LIBRARY\_PATH} variables. The HPE Cray MPI libraries (C and Fortran versions) paths are actually provided by the module \texttt{cray-mpich-abi}. 
For convenience, LUST provides the module \texttt{singularity-bindings}~\footnote{\url{https://lumi-supercomputer.github.io/LUMI-EasyBuild-docs/s/singularity-bindings/}.} that users can load to set the host environment.

\begin{figure}[t]
\begin{lstlisting}[style=SingularityStyle]
Bootstrap: localimage
From: mpich-ubuntu24.04.sif

%files
    mpitest.c /container/test_mpi/
    intercept.c /container/test_mpi/

%environment
    export LD_PRELOAD="/container/test_mpi/intercept.so"

%post -c /bin/bash
    apt-get update && apt-get -y upgrade --no-install-recommends

    # Cleanup
    apt-get autoremove -y
    apt-get clean && rm -rf /var/lib/apt/lists/*

    cd /container/test_mpi/

    # Compile mpitest
    mpicc -o mpitest.x mpitest.c

    # Compile the intercept library
    mpicc -shared -fPIC -ldl -o intercept.so intercept.c

    # Build OSU benchmarks
    OSU_NAME=osu-micro-benchmarks-7.4
    wget -q http://mvapich.cse.ohio-state.edu/download/mvapich/${OSU_NAME}.tar.gz
    tar xf ${OSU_NAME}.tar.gz
    cd ${OSU_NAME}
    ./configure --prefix=/container/osu CC=$(which mpicc) CXX=$(which mpicxx) 
    make -j$(getconf _NPROCESSORS_ONLN) install
    cd ..
    rm -rf ${OSU_NAME} && rm ${OSU_NAME}.tar.gz
\end{lstlisting}
\caption{Singularity definition file, based on MPICH base image, used to test if MPI within the container is properly running (via the \texttt{mpitest.x} and the \texttt{intercept.so}) and its performance (OSU Micro Benchmarks).
}
\label{fig:mpich_app}
\end{figure}

We run some tests on LUMI to check that the container based on MPICH works as expected. In the following \texttt{srun} commands, we omit the account (\texttt{-A}) and the partition (\texttt{-p}) flags.
\begin{itemize}
    \item Check which MPI library is used at runtime:\\
\begin{minipage}{\linewidth}
\begin{lstlisting}[style=TerminalStyle]
$ srun -n 1 singularity exec mpich-test.sif /container/test_mpi/mpitest.x
# ranks = 1 (rank id = 0)

MPI VERSION    : CRAY MPICH version 8.1.27.26 (ANL base 3.4a2)
MPI BUILD INFO : Fri Aug 11  0:25 2023 (git hash 55e934a)

\end{lstlisting}
\end{minipage}\\
 HPE Cray MPI is correctly recognized.
    \item Report an error (by setting \texttt{CHECK\_MPI=1}) if the module \linebreak \texttt{singularity-bindings} is not loaded:
    \\
\begin{minipage}{\linewidth}
\begin{lstlisting}[style=TerminalStyle]
$ srun -n 2 singularity exec mpich-test.sif /container/test_mpi/mpitest.x
# ranks = 1 (rank id = 0)

MPICH Version:	3.4a2
MPICH Release date:	Wed Dec 18 14:46:35 CST 2019
MPICH ABI:	12:0:0
MPICH Device:	ch3:nemesis
MPICH configure:	--prefix=/container/mpi --disable-static --disable-rpath --disable-wrapper-rpath --enable-fast=all,O3 --with-device=ch3 --mandir=/usr/share/man
MPICH CC:	gcc    -DNDEBUG -DNVALGRIND -O3
MPICH CXX:	g++   -DNDEBUG -DNVALGRIND -O3
MPICH F77:	gfortran -fallow-argument-mismatch  -O3
MPICH FC:	gfortran   -O3

# ranks = 1 (rank id = 0)

MPICH Version:	3.4a2
MPICH Release date:	Wed Dec 18 14:46:35 CST 2019
MPICH ABI:	12:0:0
MPICH Device:	ch3:nemesis
MPICH configure:	--prefix=/container/mpi --disable-static --disable-rpath --disable-wrapper-rpath --enable-fast=all,O3 --with-device=ch3 --mandir=/usr/share/man
MPICH CC:	gcc    -DNDEBUG -DNVALGRIND -O3
MPICH CXX:	g++   -DNDEBUG -DNVALGRIND -O3
MPICH F77:	gfortran -fallow-argument-mismatch  -O3
MPICH FC:	gfortran   -O3

$ CHECK_MPI=1 srun -n 2 singularity exec mpich-test.sif /container/test_mpi/mpitest.x
ERROR: # MPI ranks 1 != # Slurm tasks 2 (rank id = 0)
application called MPI_Abort(MPI_COMM_WORLD, 1) - process 0
[unset]: write_line error; fd=-1 buf=:cmd=abort exitcode=1
:
system msg for write_line failure : Bad file descriptor
ERROR: # MPI ranks 1 != # Slurm tasks 2 (rank id = 0)
application called MPI_Abort(MPI_COMM_WORLD, 1) - process 0
[unset]: write_line error; fd=-1 buf=:cmd=abort exitcode=1
:
system msg for write_line failure : Bad file descriptor
srun: error: nid005347: tasks 0-1: Exited with exit code 1
srun: launch/slurm: _step_signal: Terminating StepId=7046274.0
\end{lstlisting}
\end{minipage}\\
Without \texttt{CHECK\_MPI=1} (by default \texttt{CHECK\_MPI=0}), we see that two sequential instances  of the application are executed (see the two corresponding outputs) and the container MPICH is used.

\item Among the OSU performance benchmarks, we report here the results of the \texttt{pt2pt} bandwidth test, when running two ranks on different nodes:
    \\
\begin{minipage}{\linewidth}
\begin{lstlisting}[style=TerminalStyle]
$ CHECK_MPI=1 srun -n 2 --ntasks-per-node=1 singularity exec mpich-test.sif \ /container/osu/libexec/osu-micro-benchmarks/mpi/pt2pt/osu_bw

# OSU MPI Bandwidth Test v7.4
# Datatype: MPI_CHAR.
# Size      Bandwidth (MB/s)
1                       2.01
2                       4.03
4                       8.05
8                      16.13
16                     32.20
32                     64.39
64                    128.81
128                   257.57
256                   490.31
512                   976.65
1024                 1949.08
2048                 3890.51
4096                 7763.86
8192                14551.28
16384               17642.47
32768               19151.75
65536               20953.49
131072              21800.11
262144              22206.10
524288              22370.86
1048576             22475.48
2097152             22527.21
4194304             22553.94
\end{lstlisting}
\end{minipage}\\
These values are in agreement with the corresponding results obtained on LUMI without the container. Same conclusion has been found for the other OSU benchmarks. 
\end{itemize}

We can adapt the Singularity definition file reported in Fig.~\ref{fig:mpitest} to build a container based on the OpenMPI base image, by changing the corresponding base image in the \texttt{From:} section. Furthermore, since we run this container under \texttt{mpixlate}, we cannot define the \texttt{LD\_PRELOAD} intercept library into the \texttt{\%environment} section to avoid conflict with the tool. We replace this section with a \texttt{\%runscript} section:\\
\begin{minipage}{\linewidth}
\begin{lstlisting}[style=SingularityStyle]
%runscript
    export LD_PRELOAD="${LD_PRELOAD}${LD_PRELOAD:+:}/container/test_mpi/intercept.so"

    if test $# -eq 0 || test -z "$@" ; then
        bash -norc
    else
        sh -c "$@"
    fi
\end{lstlisting}
\end{minipage}\\
Then, we load the modules (in this order): \texttt{singularity-bindings}, \linebreak
\texttt{cray-mpich}, and \texttt{cray-mpixlate}. We need to prepend the \texttt{mpixlate} libraries path to \texttt{SINGULARITYENV\_LD\_LIBRARY\_PATH}. Finally, we can run some tests, similar to what we did for the MPICH container, where we specify \texttt{mpixlate -s ompi.40 -t cmpich.12} in the Slurm launch command (see example below). 

\begin{itemize}
    \item Check which MPI library is used at runtime:\\
\begin{minipage}{\linewidth}
\begin{lstlisting}[style=TerminalStyle]
$ srun -n 1 mpixlate -s ompi.40 -t cmpich.12 singularity run openmpi-test.sif \ 
       /container/test_mpi/mpitest.x
# ranks = 1 (rank id = 0)

MPI VERSION    : CRAY MPICH version 8.1.27.26 (ANL base 3.4a2)
MPI BUILD INFO : Fri Aug 11  0:25 2023 (git hash 55e934a)

\end{lstlisting}
\end{minipage}\\
 HPE Cray MPI is correctly recognized. We can check that OpenMPI is used inside the container by simply running:\\
\begin{minipage}{\linewidth}
\begin{lstlisting}[style=TerminalStyle]
$ singularity run openmpi-test.sif /container/test_mpi/mpitest.x
# ranks = 1 (rank id = 0)

Open MPI v4.1.6, package: Debian OpenMPI, ident: 4.1.6, repo rev: v4.1.6, Sep 30, 2023
\end{lstlisting}
 \end{minipage}\\
 
\item We tested the OSU benchmarks, which are reporting the same performance obtained for the MPICH container, e.g.:\\
\begin{minipage}{\linewidth}
\begin{lstlisting}[style=TerminalStyle]
$ CHECK_MPI=2 srun -n 2 --ntasks-per-node=1 -A project_462000031 -p standard-g \
    mpixlate -s ompi.40 -t cmpich.12 singularity run openmpi-test.sif \
    /container/osu/libexec/osu-micro-benchmarks/mpi/pt2pt/osu_bw
INFO: # MPI ranks 2 == # Slurm tasks 2 (rank id = 0)
INFO: # MPI ranks 2 == # Slurm tasks 2 (rank id = 1)

# OSU MPI Bandwidth Test v7.4
# Datatype: MPI_CHAR.
# Size      Bandwidth (MB/s)
1                       1.97
2                       3.94
4                       7.86
8                      15.76
16                     31.58
32                     63.13
64                    126.20
128                   252.03
256                   480.91
512                   961.12
1024                 1919.49
2048                 3840.13
4096                 7652.95
8192                14536.25
16384               17704.78
32768               19263.25
65536               21124.01
131072              21829.02
262144              22223.10
524288              22423.76
1048576             22514.62
2097152             22556.77
4194304             22579.80

\end{lstlisting}
 \end{minipage}

\end{itemize}

\section{Conclusion}
In this paper, we have presented how users can deploy containerized applications on the LUMI supercomputer via the Slurm workload manager, specifically for running applications based on MPI parallelization. We have shown how users can build and run containers on LUMI based on the two main MPI implementations, namely MPICH and OpenMPI. We have presented some examples to show their functionalities and performance, including some tests to check that they properly run. The proposed MPI containers can be provided on LUMI so that users can use them as base images, either deriving from the existing container or re-building from scratch. Similar concepts can be applied to the case of other supercomputers.

As future work, we aim to explore: the OFI Libfabric injection proposed in Ref.~\cite{cscs_libfabric}, testing of other MPI implementations (e.g. MVAPICH), the OpenMPI solution proposed in Ref.~\cite{openmpi_frontier}, and the possibility to include GPU-aware MPI support.

\begin{credits}
\subsubsection{\ackname} The work presented in this article has been carried out as part of LUMI HPE Centre of Excellence activities. The author would like to thank the LUMI User Support Team members. A special thank to Tiziano M\"uller and Harvey Richardson of the HPE HPC/AI EMEA Research Lab for the useful discussions and comments.

\subsubsection{\discintname}
The authors have no competing interests to declare that are relevant to the content of this article.
\end{credits}

%
%
%
\bibliographystyle{splncs04}
\bibliography{paper}

\begin{thebibliography}{10}
\providecommand{\url}[1]{\texttt{#1}}
\providecommand{\urlprefix}{URL }
\providecommand{\doi}[1]{https://doi.org/#1}

\bibitem{openmpi}
{Open MPI: Open Source High Performance Computing},
  \url{https://www.open-mpi.org/}

\bibitem{cve_userns}
{CVE-2022-0185} (2022), \url{https://nvd.nist.gov/vuln/detail/CVE-2022-0185}

\bibitem{mpich}
{Argonne National Laboratory}: {MPICH: A High-Performance, Portable
  Implementation of MPI},
  \url{https://www.anl.gov/mcs/mpich-a-highperformance-portable-implementation-of-mpi}

\bibitem{sarus}
Benedičič, L., Cruz, F., Madonna, A., Mariotti, K.: {Sarus: Highly Scalable
  Docker Containers for HPC Systems}, pp. 46--60 (12 2019).
  \doi{10.1007/978-3-030-34356-9_5}

\bibitem{slingshot}
De~Sensi, D., Di~Girolamo, S., McMahon, K.H., Roweth, D., Hoefler, T.: {An
  In-Depth Analysis of the Slingshot Interconnect}. In: {SC20: International
  Conference for High Performance Computing, Networking, Storage and Analysis}.
  pp. 1--14 (2020). \doi{10.1109/SC41405.2020.00039}

\bibitem{osti_1886029}
Dykstra, D.: {Apptainer Without Setuid}  (8 2022),
  \url{https://www.osti.gov/biblio/1886029}

\bibitem{securityhpc}
Gantikow, H., Reich, C., Knahl, M., Clarke, N.: {Providing Security in
  Container-Based HPC Runtime Environments}. vol.~9945 (06 2016).
  \doi{10.1007/978-3-319-46079-6_48}

\bibitem{gerhardt2017shifter}
Gerhardt, L., Bhimji, W., Fasel, M., Porter, J., Mustafa, M., Jacobsen, D.,
  Tsulaia, V., Canon, S.: {Shifter: Containers for HPC}. In: {J. Phys. Conf.
  Ser.} vol.~898, p. 082021 (2017)

\bibitem{libfabric}
Grun, P., Hefty, S., Sur, S., Goodell, D., Russell, R.D., Pritchard, H.,
  Squyres, J.M.: {A Brief Introduction to the OpenFabrics Interfaces - A New
  Network API for Maximizing High Performance Application Efficiency}. In:
  {2015 IEEE 23rd Annual Symposium on High-Performance Interconnects}. pp.
  34--39 (2015). \doi{10.1109/HOTI.2015.19}

\bibitem{heon_2021_4735634}
Heon, M., Walsh, D., Baude, B., Mohnani, U., Cui, A., Sweeney, T., Scrivano,
  G., Evich, C., Rothberg, V., Trmač, M., Honce, J., Wang, Q., Mandvekar, L.,
  Reber, A., Santiago, E., Grunert, S., Dahyabhai, N., Bjorklund, A., Kushwaha,
  K., Ashwin~Sha, S., Pu, Y., Zhangguanzhang, Vasek, M., Community, P.:
  {Podman: A tool for managing OCI containers and pods} (May 2021).
  \doi{10.5281/zenodo.4735634}, \url{https://doi.org/10.5281/zenodo.4735634}

\bibitem{craympich}
Kandalla, K., McMahon, K., Ravi, N., White, T., Kaplan, L., Pagel, M.:
  {Designing the HPE Cray Message Passing Toolkit Software Stack for HPE Cray
  EX Supercomputers}. In: {CUG2023 Proceedings} (May 2023),
  \url{https://cug.org/proceedings/cug2023_proceedings/includes/files/pap144s2-file1.pdf}

\bibitem{kurtzer2017singularity}
Kurtzer, G.M., Sochat, V., Bauer, M.W.: {Singularity: Scientific containers for
  mobility of compute}. PloS one  \textbf{12}(5) (2017)

\bibitem{cscs_libfabric}
Madonna, A., Aliaga, T.: {Libfabric-based Injection Solutions for Portable
  Containerized MPI Applications}. In: {2022 IEEE/ACM 4th International
  Workshop on Containers and New Orchestration Paradigms for Isolated
  Environments in HPC (CANOPIE-HPC)}. pp. 45--56 (2022).
  \doi{10.1109/CANOPIE-HPC56864.2022.00010}

\bibitem{merkel2014docker}
Merkel, D.: {Docker: lightweight linux containers for consistent development
  and deployment}. Linux journal  \textbf{2014}(239), ~2 (2014)

\bibitem{charliecloud_seccomp}
Phinney, M.: {New Root Emulation Mode for Charliecloud Using seccomp}. In:
  {2023 IEEE/ACM International Workshop on Containers and New Orchestration
  Paradigms for Isolated Environments in HPC (CANOPIE-HPC)} (2023)

\bibitem{fakeroot}
Priedhorsky, R., Canon, R.S., Randles, T., Younge, A.J.: {Minimizing privilege
  for building HPC containers}. In: {Proceedings of the International
  Conference for High Performance Computing, Networking, Storage and Analysis}.
  SC '21, Association for Computing Machinery, New York, NY, USA (2021).
  \doi{10.1145/3458817.3476187}, \url{https://doi.org/10.1145/3458817.3476187}

\bibitem{priedhorsky2017charliecloud}
Priedhorsky, R., Randles, T.: {Charliecloud: Unprivileged containers for
  user-defined software stacks in hpc}. In: {Proceedings of the International
  Conference for High Performance Computing, Networking, Storage and Analysis}.
  pp. 1--10 (2017)

\bibitem{openmpi_frontier}
Pritchard, H., Naughton, T., Shehata, A., Bernholdt, D.: {Open MPI for HPE Cray
  EX Systems}. In: {CUG2023 Proceedings} (May 2023),
  \url{https://cug.org/proceedings/cug2023_proceedings/includes/files/pap140s2-file1.pdf}

\bibitem{hybrid_mode}
Sande~Veiga, V., Simon, M., Azab, A., Fernandez, C., Muscianisi, G., Fiameni,
  G., Marocchi, S.: {Evaluation and Benchmarking of Singularity MPI containers
  on EU Research e-Infrastructure}. In: {2019 IEEE/ACM International Workshop
  on Containers and New Orchestration Paradigms for Isolated Environments in
  HPC (CANOPIE-HPC)}. pp. 1--10 (2019).
  \doi{10.1109/CANOPIE-HPC49598.2019.00006}

\end{thebibliography}
\end{document}